# Modified geometrical optics of a smoothly inhomogeneous isotropic medium: the anisotropy, Berry phase, and the optical Magnus effect

K.Yu. Bliokh[1*], Yu.P. Bliokh[2]

[1]*Institute of Radio Astronomy, 4 Krasnoznamyonnaya st., Kharkov, 61002, Ukraine*
[2]*Department of Physics, Technion, Haifa, 32000, Israel*

In this paper we present a modification of the geometrical optics method, which allows one to properly separate the complex amplitude and the phase of the wave solution. Appling this modification to a smoothly inhomogeneous isotropic medium, we show that in the first geometrical optics approximation the medium is weakly anisotropic. The refractive index, being dependent on the direction of the wave vector, contains the correction, which is proportional to the Berry geometric phase. Two independent eigenmodes of right-hand and left-hand circular polarizations exist in the medium. Their group velocities and phase velocities differ. The difference in the group velocities results in the shift of the rays of different polarizations (the optical Magnus effect). The difference in the phase velocities causes the increase of Berry phase along with the interference of two modes leading to the familiar Rytov law about the rotation of the polarization plane of a wave. The theory developed suggests that both the optical Magnus effect and the Berry phase are accompanying nonlocal topological effects. In this paper the Hamilton ray equations giving a unified description for both of these phenomena have been derived and also a novel splitting effect for a ray of noncircular polarization has been predicted. Specific examples are also discussed.

---

[*] E-mail: kostya@bliokh.kharkiv.com, k_bliokh@mail.ru

PACS numbers: 41.20.Jb, 42.15.-i, 41.85.Ct

## 1. INTRODUCTION

The first consistent presentation of the geometrical optics approximation, as applied to the electromagnetic wave propagation through a smoothly inhomogeneous isotropic medium, was given by Rytov in [1]. There was indicated that in the zero geometrical optics approximation, only the phase and the amplitude of a transverse wave can be determined, but not the polarization. This is due to the fact that two modes with distinct polarizations turn out to be degenerate or indistinguishable. The polarization degeneracy can be removed through a consideration of the first-order terms in the geometrical optics approximation. Hence the familiar Rytov law about rotation of the polarization plane of an electromagnetic wave in a smoothly inhomogeneous medium follows [1–3]. The geometrical properties of this law were detailed by Vladimirsky in [2]. Subsequently, it was shown that the Rytov law is nothing but a consequence of the appearance of the Berry geometric phases of photons (see [4−6]).

The anisotropic medium differs from the isotropic one in that (in the general case) it has no polarization degeneracy and thus the polarization of electromagnetic waves is determined even in zero geometrical optics approximation [3]. In this regard, the account of the first geometrical optics approximation in an isotropic medium is similar to the case of a weakly anisotropic medium. Provided this analogy has good grounds, what this means is the smooth inhomogeneity causes a real weak anisotropy of the medium. In this case, the assumed anisotropy will result in the propagation of the eigenmodes (waves of right-hand and left-hand polarizations) along different trajectories.

The changes of ray trajectories with polarization correspond to the so-called optical Magnus effect, which was suggested in 1990 by Zel'dovich and coauthors [7]. The optical Magnus effect was calculated theoretically and supported experimentally for waves in optical fibers. After that the phenomenological theory describing this phenomenon in the geometrical optics approximation was advanced in [8]. The results of the present work support and generalize the corrections introduced by Liberman and Zel'dovich and demonstrate that the relevant equations and effects follow from the initial principles of geometrical optics.

Below is shown that in the first (Rytov) geometrical optics approximation, an isotropic smoothly inhomogeneous medium *is actually anisotropic*. What this means is: 1) the refractive index of this medium depends on the wave-vector direction; 2) the medium contains two independent transverse modes with right-hand and left-hand polarizations, their group velocities and phase velocities are distinct; 3) as a consequence of the anisotropy, the right-hand polarized and left-hand polarized modes propagate along different ray trajectories.

In that way a ray of the wave with mixed (not circular) polarization is split into two independent rays with right-hand and left-hand polarizations. This fact makes a prediction about a novel phenomenon, which is not covered by the theory of the optical Magnus effect [7,8]. Really, Zel'dovich theory describes the *displacement* of the ray's center of gravity depending on its polarization, but does not point to a possible ray *splitting*. Our theory suggests that only circular polarized independent rays exist in the framework of the approximation considered. The rays of other polarizations arise from the interference of the eigenmodes that propagate along different trajectories.

In fact, the theory developed establishes a link between two fundamental phenomena – the Berry geometrical phase and the optical Magnus effect. It is shown in the paper that the former implies the difference of phase velocities of the eigenmodes, whereas the latter is caused by the difference of group velocities. We demonstrate that the optical Magnus effect, as well as the Berry phase, is a nonlocal topological effect described by the geometry of the system's trajectory in a momentum space.

The results mentioned follow immediately from the initial principles of geometrical optics. The reason why these phenomena have not been theoretically revealed before is that in the conventional geometrical optics (see, for example, [3]) the separation of the complex amplitude and the complex phase was performed not quite correctly. As a result, the terms of the first order, which cause the above-mentioned effects, have been disregarded in the wave eikonal. Below the modified geometrical optics theory is presented, which is free of the drawbacks outlined. Specific examples are also analyzed (Section 4).

## 2. SEPARATION OF THE PHASE AND THE AMPLITUDE IN THE WAVE SOLUTIONS

In the conventional geometrical optics, the monochromatic wave field in an isotropic medium can be written as

$$\mathbf{E} \approx \left(\mathbf{E}^{(0)} + k_0^{-1}\mathbf{E}^{(1)} + k_0^{-2}\mathbf{E}^{(2)} + ...\right)\exp(ik_0\psi) , \quad (1)$$

where $k_0^{-1} = c/\omega$ is the inverse wave number in vacuum, which is small as compared to the typical scale of inhomogeneity. The phase $\psi$ can be determined from the eikonal equation

$$(\nabla\psi)^2 = n^2 \equiv \varepsilon \quad (2)$$

and $\frac{\partial\psi}{\partial t} = -c$, while the amplitudes $\mathbf{E}^{(k)}$ are found from the transport equations of the relevant order. It is assumed that the phase characteristics of wave are determined by the phase $\psi$ and the eikonal equation, whereas the amplitudes $\mathbf{E}^{(k)}$ and the transport equations specify the current amplitude of the wave and its polarization.

However, in general, this is not the case. Actually, the adiabatic (or WKB, or geometrical optics) solution to the stationary wave equation is constructed by the perturbation method for the unitary complex phase in the exponent (see, [9−11]):

$$\mathbf{E} \approx \exp(ik_0 \hat{\Phi}) \mathbf{E}_0 \equiv \exp\left( ik_0 \sum_{k=0}^{m} k_0^{-k} \hat{\Psi}^{(k)} \right) \mathbf{E}_0 \ . \tag{3}$$

Here $m = 0, 1, 2, ...$ is the approximation order, $\mathbf{E}_0$ is the field initial value, while $\hat{\Phi}$ and $\hat{\Psi}^{(k)}$ are the matrix operators since the field is a vector changing its direction. Some of the terms comprising complex phase of Eq. (3) can be taken out and inserted into the pre-exponent factor (amplitude). It is obvious that the separation of these terms into the phase and the amplitude is a matter of convention (in so far as the amplitude is a complex value). Hence, it is primarily important to define a criterion, according to which we can separate these terms.

In the conventional geometrical optics [3] the phase $\psi$ and the eikonal equation correspond to zero order approximation in (3):

$$\psi = \hat{\Psi}^{(0)} \equiv \Psi^{(0)} \tag{4}$$

($\hat{\Psi}^{(0)}$ is a scalar, or a diagonal operator with the equal eigenvalues; this is just the polarization degeneracy). The amplitude $\mathbf{E}^{(k)}$ and the associated transport equation correspond to perturbations of order $(k+1)$:

$$\mathbf{E}^{(0)} = \exp(i\hat{\Psi}^{(1)}) \mathbf{E}_0 \ , \text{ and so on.}^{1)} \tag{5}$$

In this equation, $\hat{\Psi}^{(1)}$ is now the operator with different eigenvalues, which determines the Rytov evolution of wave polarization. Thus, in the framework of conventional geometrical optics, the phase and the amplitude are separated according to their orders, for which, actually, there are no grounds.

---

[1] We assume that the operator in the exponent in (3) can be put in a diagonal form, corresponding to a basis of normal independent modes. Hence, by reason of asymptotic nature of the series, all operators under the summation sign will be diagonal in this basis. By this is meant that these operators commute, since from here on we shall be able to represent the exponent of their sum as a product of the exponents.

We suggest another way. Note that the phase is a *nonlocal* or *integral* value, since its increment is determined by the entire path covered by the wave. To the contrary, the amplitude in a passive nonabsorbing medium with independent eigenmodes is conceptually a *local* value, being dependent only on the initial conditions and the current values of parameters. Indeed, the amplitude specifies the wave energy, whose variation is bound to be governed by the initial and final points only, and not by the transfer path.[2] In the case of an absorbing or active medium, the amplitude is no longer a local value. Then, the local amplitude should be multiplied by the part of nonlocal exponential function with a real exponent.

Thus the procedure for separating the phase and the amplitude is as follows. In Eq. (3) we separate local and nonlocal terms:

$$\mathbf{E} \approx \exp\left(ik_0\hat{\Phi}^{(loc)} + ik_0\hat{\Phi}^{(nonloc)}\right)\mathbf{E}_0 \ . \tag{6}$$

Then, the amplitude and the phase can be separated in the following manner:

$$\mathbf{E} \approx \hat{\mathbf{A}} \exp\left(i\hat{\phi}\right)\mathbf{E}_0 \ , \tag{7}$$

where

$$\hat{\mathbf{A}} = \exp\left(ik_0\hat{\Phi}^{(loc)}\right), \ \hat{\phi} = k_0\hat{\Phi}^{(nonloc)} \ . \tag{8}$$

The eigenvectors of the operator $\exp\left(ik_0\hat{\Phi}\right)$ determine the medium eigenmodes at each point. At that the derivatives $\dfrac{\partial}{\partial t}$ and $\dfrac{\partial}{\partial \mathbf{r}}$ of the eigenvalues of the operator $k_0\hat{\Phi}^{(nonloc)}$ determine their complex frequencies and wave-vectors of the medium's independent modes, while the eigenvectors of the operator $\hat{\mathbf{A}}$ specify the wave polarization. It should be noted that the separation of the values into local and nonlocal ones is ambiguous and is determined up to the gauge transformation

$$\hat{\phi} \rightarrow \hat{\phi} + \hat{\varphi} \ , \ \hat{\mathbf{A}} \rightarrow \hat{\mathbf{A}} \exp\left(-i\hat{\varphi}\right) \ , \tag{9}$$

---

[2] Probably, the existence condition for an adiabatic invariant serves as the locality condition for the amplitude (see [3,12]). In [10] it is proved for linear ordinary differential equations.

where $\hat{\varphi}$ is the local scalar potential. However, as will be seen from the next section, these transformations have no effect on the physically observable values.

## 3. GEOMETRICAL OPTICS OF A SMOOTHLY INHOMOGENEOUS ISOTROPIC MEDIUM

**3.1. Eikonals and refractive indices.** In order to derive correct characteristics of a smoothly inhomogeneous isotropic medium, let us use the familiar formulas for the wave eikonals of the right-hand and left-hand circularly polarized waves. They follow immediately from Maxwell's equations and can be given as [1,6]

$$\phi^\pm = \int_0^s k^{(0)} ds \pm \vartheta \ . \tag{10}$$

Here $k^{(0)} = n^{(0)}(\mathbf{r}) k_0$ stands for the current wave number, $n^{(0)}(\mathbf{r}) = \sqrt{\varepsilon(\mathbf{r})}$ is the local refractive index of the relevant isotropic medium, $k_0 = \omega/c$, $s$ is the length of the ray arc, and $\vartheta$ is Berry geometric phase, which has opposite signs for the waves of right-hand and left-hand polarizations. We have assumed in Eq. (10) that $\phi^\pm\big|_{s=0} = 0$, since any constant additions can be included into complex amplitudes and bellow we will use only gradients of the eikonals (10). The superscript (0) indicates that the current values correspond to the zero-order geometrical optics approximation. Below we will derive the corrections to the wave vectors and to the refractive indices. Here and further the medium smoothness implies the short-wave asymptotic $k_0 \equiv \omega/c \to \infty$, whereas formula (10) is derived in the first approximation in $k_0^{-1}$. The first-order correction terms are contained in the Berry phase, which can be given in the form [6]

$$\vartheta = \int_0^s \mathbf{G}\dot{\mathbf{p}} ds = \int_L \mathbf{G} d\mathbf{p} \ . \tag{11}$$

Here, the dimensionless wave momentum $\mathbf{p} = \mathbf{k}/k_0$ has been introduced, $\mathbf{G} = \mathbf{G}(\mathbf{p})$ is a certain nonpotential field in the $\mathbf{p}$-space, the dot signifies the differentiation with respect to $s$ (that is, along the ray), and $L$ is the contour along which the system is moving in the $\mathbf{p}$-space. Equation (11), as well as all first-order correction terms below, is calculated along the trajectories of the zero-order approximation (i.e. with $\mathbf{p} = \mathbf{p}^{(0)} = \mathbf{k}^{(0)}/k_0$). The field $\mathbf{G}$ is not uniquely defined; particularly, it can be chosen in the form $\mathbf{G}(\mathbf{p}) = \dfrac{(\mathbf{p}\mathbf{a})[\mathbf{p}\times\mathbf{a}]}{p[\mathbf{p}\times\mathbf{a}]^2}$, where $p = |\mathbf{p}|$ and $\mathbf{a}$ is an arbitrary constant vector. With gauge transformations (9), the field $\mathbf{G}$ transforms as

$$\mathbf{G} \to \mathbf{G} + \frac{\partial \varphi}{\partial \mathbf{p}}, \qquad (12)$$

(where $\varphi(\mathbf{p})$ is an arbitrary scalar potential), while the physically measurable values are related to the curl of the field $\mathbf{G}$, which is equal to [3]

$$\left[\frac{\partial}{\partial \mathbf{p}} \times \mathbf{G}\right] = -\frac{\mathbf{p}}{p^3}. \qquad (13)$$

Eq. (13) is obtained by explicit differentiation of $\mathbf{G}$ and determines the gauge-invariant magnetic monopole type structure in the waves' momentum space (see [17]). Geometric phase (11) may be considered as an integral along the ray as well as a contour integral in the $\mathbf{p}$-space.

Let us determine the refractive indices of the right and left circularly polarized waves by writing the eikonal equation for Eqs. (10), (11) as

$$n^{\pm} = k_0^{-1}|\nabla \phi^{\pm}| = n^{(0)}(\mathbf{r}) + n^{(1)}(\mathbf{r}, \mathbf{p}), \qquad n^{(1)} = \pm k_0^{-1}\frac{d\vartheta}{ds} = \pm k_0^{-1}\mathbf{G}\dot{\mathbf{p}}. \qquad (14)$$

In the conventional geometrical optics [1,3], the correction term $n^{(1)}$ did not arise [13] since the eikonal was derived in zero-order approximation in $k_0^{-1}$, while all the higher-order terms (including the nonlocal factors $\exp(\pm i\vartheta)$ associated with the geometric phase) pertained to the transport equation, i.e. to the amplitude. Meanwhile, the geometric phase is a nonlocal value, which cannot be attributed to the wave amplitude. The Berry phase can distort substantially the phase front. For example, for a ray with torsion, the phase front gradient has an additional fixed component $\pm\nabla\vartheta$, which changes the wave vector and the phase velocity. As will be seen, the obtained correction term $n^{(1)}$ leads to the corrections in the geometrical optics equations, which are supported experimentally, and hence, have real physical grounds. Note that in view of the essential dependence of the geometric phase on the ray trajectory, the correction term $n^{(1)}$ depends not only on the current coordinate $\mathbf{r}$, but also on the wave vector *direction*, that is, on the wave *momentum*. This points to the weak anisotropy of a locally isotropic medium.

**3.2. Basic equations.** Let us write the Hamiltonian equations for ray propagation [3]. By choosing the Hamiltonian as $H = p - n^{\pm}(\mathbf{r}, \mathbf{p}) = 0$ and using Eqs. (14), we have

$$\frac{d\mathbf{p}}{ds} = \frac{\partial n^{\pm}}{\partial \mathbf{r}} = \frac{\partial n^{(0)}}{\partial \mathbf{r}} \pm k_0^{-1} \frac{\partial}{\partial \mathbf{r}} \left( \frac{d\vartheta}{ds} \right) = \frac{\partial n^{(0)}}{\partial \mathbf{r}} \pm k_0^{-1} \frac{\partial}{\partial \mathbf{r}} (\mathbf{G}\dot{\mathbf{p}}) , \qquad (15)$$

$$\frac{d\mathbf{r}}{ds} = -\frac{\partial n^{\pm}}{\partial \mathbf{p}} = \mathbf{l} \mp k_0^{-1} \frac{\partial}{\partial \mathbf{p}} \left( \frac{d\vartheta}{ds} \right) = \mathbf{l} \mp k_0^{-1} \frac{\partial}{\partial \mathbf{p}} (\mathbf{G}\dot{\mathbf{p}}) . \qquad (16)$$

Here $\mathbf{l} = \mathbf{p}/p$ is the unit vector of the normal to the wave phase front. (At the same time, it is the unit tangent vector of the ray in zero approximation in $k_0^{-1}$.) It worth noticing also, that $\dot{\mathbf{p}}$ term (see Eq. (14)) in the ray Hamiltonian should be interpreted not as independent quantity but only as expressed in the end from the zero order Hamiltonian equations (see bellow).

Equations (15) and (16) can be analyzed by applying the perturbation method in $k_0^{-1}$. By representing all values in the form $a = a^{(0)} + a^{(1)}$ ($a^{(0)} \sim 1$, $a^{(1)} \sim k_0^{-1}$), we have from Eqs. (15) and (16) in zero approximation

$$\frac{d\mathbf{p}^{(0)}}{ds} = \frac{\partial n^{(0)}}{\partial \mathbf{r}} , \quad \frac{d\mathbf{r}^{(0)}}{ds} = \mathbf{l}^{(0)} . \qquad (17)$$

These are the familiar geometrical optics equations for an isotropic medium [3]. The second terms in the right sides of Eqs. (15) and (16) introduce the corrections of the order of $k_0^{-1}$, and hence they should be considered on the solutions (trajectories) of zero approximation. As a result, for the first-order corrections we obtain

$$\frac{d\mathbf{p}^{(1)}}{ds} = \pm k_0^{-1} \frac{\partial}{\partial \mathbf{r}} (\mathbf{G}\dot{\mathbf{p}})^{(0)} , \quad \frac{d\mathbf{r}^{(1)}}{ds} = \mp k_0^{-1} \frac{\partial}{\partial \mathbf{p}} (\mathbf{G}\dot{\mathbf{p}})^{(0)} , \qquad (18)$$

where the superscript (0) signifies that all values in the right-hand sides of (18) are derived from zero-order equations (17). Here and further it is considered that $\mathbf{l}^{(1)} = 0$ and $\mathbf{l} = \mathbf{l}^{(0)}$. As we will see later, equations (18) describe the deviations in a wave momentum and coordinates that are associated with the spatial and momentum gradients of the Berry phase, respectively. As will be seen, the first equation in (18) governs the emergence of Berry phase, while the second equation describes the deviations of the rays of different polarizations by virtue of the optical Magnus effect [7].

It is significant that the wave evolution for right and left circular polarizations is given by independent equations and thus these waves are the independent medium eigenmodes. This fact correlates well with the quantum-mechanical notion of photons, according to which a photon may possess the helicity equal to +1 or −1 only, which correspond to right and left circular polarizations. In the framework of a given approximation, an arbitrarily polarized wave cannot be treated independently, but only as a superposition of circular eigenmodes.

**3.3. Equation for momentum, Berry phases, and phase velocities.** Consider initially the first equation in (18). First of all, let us note that after integration with the operator $k_0 \int d\mathbf{r} \int dt$, it exactly defines the geometrical term $\vartheta$ in the phase (10), (11). The first equation (18) is responsible for the change of the *momentum* (wave vector) and the *phase* velocity of waves in *absolute value*, but not direction. To prove this, let's multiply scalarly the first equation (18) by $\mathbf{l}$ and, taking into account that $\mathbf{l}\partial/\partial \mathbf{r} = d/ds$, we obtain

$$\frac{dp^{(1)}}{ds} = \mathbf{l}\frac{d\mathbf{p}^{(1)}}{ds} = \pm k_0^{-1} \frac{d}{ds}(\mathbf{G}\dot{\mathbf{p}})^{(0)} , \qquad (19)$$

Consequently, in the first geometrical optics approximation, the wave momentum (wave vector) is

$$\mathbf{p} = \mathbf{p}^{(0)} \pm k_0^{-1}(\mathbf{G}\dot{\mathbf{p}})^{(0)}\mathbf{l} . \qquad (20)$$

When integrating Eq. (19), we assume for simplicity that $\mathbf{p}^{(1)}(0) = 0$. Equation (20) follows immediately from the initial expressions (10), (11), and (14) for eikonals and refractive indices.

When integrating along the ray, two terms in Eq. (20) represent the dynamic phase and the geometric phase (parts of the eikonal), respectively. From Eq. (20) or Eq. (14) we have the following expression for the phase velocities of the left-hand and right-hand waves

$$\mathbf{v}_{ph}^{\pm} = \frac{c}{n^{(0)}} \left(1 \mp \frac{1}{n^{(0)} k_0} \frac{d\vartheta}{ds}\right) \mathbf{l} = \frac{c}{n^{(0)}} \left(1 \mp \frac{(\mathbf{G}\dot{\mathbf{p}})^{(0)}}{n^{(0)} k_0}\right) \mathbf{l} \ . \tag{21}$$

It should be noted that the right-hand side of the first equation in (18) involves also the component that is orthogonal to $\mathbf{l}$. Nominally, it causes the deviation of the momentum from the direction of the zero momentum $\mathbf{p}^{(0)}$. However, this deviation does not exceed $k_0^{-1}$ in the order of magnitude and essentially depends on gauge transformations (9) and (12). The reason is that under the gauge transformations a certain part of the phase turns into the amplitude, with a consequent slight distortion of the phase front (or small deviations of the front normal from the zero-approximation direction). The momentum (wave vector), however, is not a physically measurable value in this range (in view of the uncertainty relation), and hence, the above-mentioned deviations are irrelevant to the values under observation. Among these values are the phase (that is, an integral of the wave vector projection onto the ray) and the ray trajectory accurate to a wavelength. From these arguments it follows that it makes sense to consider only the longitudinal component in the right-hand side of the first equation in (18) resulting in Eq. (20). After elimination of the immeasurable transversal deviations, the first equation in (18) takes the form

$$\frac{d\mathbf{p}^{(1)}}{ds} = \pm k_0^{-1} \mathbf{l} \left[\frac{\partial}{\partial \mathbf{r}} (\mathbf{G}\dot{\mathbf{p}})\right]^{(0)} = \pm k_0^{-1} \frac{d}{ds} (\mathbf{G}\dot{\mathbf{p}})^{(0)} \ . \tag{22}$$

This equation is integrable (see Eq. (20)) and, as is seen from Eq. (11), is responsible for the appearance of the Barry phase. It follows that the first-order corrections do not change the direction of the phase front normal, that is, $\mathbf{l}^{(1)} = 0$, $\mathbf{l} = \mathbf{l}^{(0)}$.

**3.4. Equation for coordinates, the optical Magnus effect, and group velocities.** We now turn our attention to the analysis of the second equation in (18). It describes the shift of the right and left circularly polarized rays, which is associated with the optical Magnus effect [7]. The right-hand side of the second equation in (18) is responsible for the ray trajectory deviations, that is, for variations in the *group* velocity. As will be seen, this correction is directed orthogonally to the ray and changes the *direction* of the group velocity. By differentiating the scalar product in the right-hand side of the second equation in (18), we obtain

$$\frac{d\mathbf{r}^{(1)}}{ds} = \mp k_0^{-1}\left[\dot{\mathbf{p}} \times \left[\frac{\partial}{\partial \mathbf{p}} \times \mathbf{G}\right]\right] \mp k_0^{-1}\left(\dot{\mathbf{p}}\frac{\partial}{\partial \mathbf{p}}\right)\mathbf{G} \; . \tag{23}$$

Here and further the superscript (0) is omitted for simplicity. Let us integrate equation (23):

$$\mathbf{r}^{(1)} = \mp k_0^{-1}\int_0^s\left\{\left[\dot{\mathbf{p}} \times \left[\frac{\partial}{\partial \mathbf{p}} \times \mathbf{G}\right]\right] + \left(\dot{\mathbf{p}}\frac{\partial}{\partial \mathbf{p}}\right)\mathbf{G}\right\}ds = \pm k_0^{-1}\int_0^s\frac{[\dot{\mathbf{p}} \times \mathbf{p}]}{p^3}ds \mp k_0^{-1}\mathbf{G}\Big|_{\mathbf{p}_0}^{\mathbf{p}} \; . \tag{24}$$

Here, formulas (13) and $\mathbf{p}_0 = \mathbf{p}(0)$ have been used.

Note now that equation (24) for the ray shift comprises two summands. The first one, being nonlocal, may grow infinitely as $s$ increases. The second summand represents a local function of the momentum $\mathbf{p}$. It cannot grow infinitely and does not exceed the wavelength $\lambda \sim k_0^{-1}$ in the order of magnitude. Evidently, the second term does not lead to observable physical effects and depends on the gauge transformations (9) and (12). This is related to the uncertainty of the notion of a ray trajectory within the range of the wavelength. Like Berry phase, the first nonlocal term in Eq. (24) is gauge-invariant. Note also that with the cyclic evolution, when the ray direction coincides with the initial one, we have $\mathbf{p} = \mathbf{p}_0$, and all nonlocal terms vanish.

Thus, when analyzing the ray shift, we have to retain only the first term in Eq. (24). As a result we have

$$\mathbf{r}^{(1)} = \pm k_0^{-1}\int_0^s\frac{[\dot{\mathbf{p}} \times \mathbf{p}]}{p^3}ds = \mp k_0^{-1}\int_L\frac{[\mathbf{p} \times d\mathbf{p}]}{p^3} \; . \tag{25}$$

The ray shift is seen to be directed orthogonally to the ray: $\mathbf{p}\dot{\mathbf{r}}^{(1)} = 0$. Formula (25) demonstrates that the ray shifts caused by the optical Magnus effect, as well as Berry geometric phase (11), can be represented as a contour integral in the **p**-space. Moreover, the shift is dictated by the geometry of the contour $L$ in the **p**-space and not by the particular $\mathbf{p}(s)$ dependence on the ray. Hence, the optical Magnus effect is a fundamental *topological* effect. The Berry phase and the Magnus effect represent wave divergences in phases and trajectories, respectively.

Displacement (25) corresponds to the differential equation that takes the place of the second equation in (18):

$$\frac{d\mathbf{r}^{(1)}}{ds} = \pm \frac{1}{k_0 p^{(0)3}} [\dot{\mathbf{p}} \times \mathbf{p}]^{(0)} . \qquad (26)$$

Equations (22) and (26) along with zero-order equations (17) describe geometrical optics of a smoothly inhomogeneous medium in the first approximation in $k_0^{-1}$. In this case, equation (12) for a momentum describes the increment of Berry phase, whereas equation (26) for a coordinate gives the shifts of differently polarized rays owing to the optical Magnus effect. By substituting the expressions $\dot{\mathbf{p}}^{(0)} = \partial n^{(0)}/\partial \mathbf{r}$ and $p^{(0)} = n^{(0)}$ from zero approximation Eqs. (17) into the right-hand sides of Eqs. (22) and (26), we obtain:

$$\frac{d\mathbf{p}^{(1)}}{ds} = \pm k_0^{-1} \mathbf{l} \left[ \frac{\partial}{\partial \mathbf{r}} \left( \mathbf{G} \frac{\partial n^{(0)}}{\partial \mathbf{r}} \right) \right] = \pm k_0^{-1} \frac{d}{ds} \left( \mathbf{G} \frac{\partial n^{(0)}}{\partial \mathbf{r}} \right) , \quad \frac{d\mathbf{r}^{(1)}}{ds} = \pm \frac{1}{k_0 n^{(0)}} \left[ \frac{\partial \ln n^{(0)}}{\partial \mathbf{r}} \times \mathbf{l} \right] . \qquad (27)$$

These 'evolutionary' equations can be solved without regard to Eqs. (17). However, the theory of Berry phases has clearly demonstrated that in a number of problems it is better to use general 'geometric' equations (22) and (26) by integrating them in the **p**-space. In particular, we could not have derived the above-discussed equations if we had not applied this approach dealing with the properties of locality and nonlocality.

Note that the second equation in (27) corresponds precisely to the correction that has been introduced into the geometrical optics equations by Zel'dovich and Liberman [8]. It has also been shown in [8] that this equation describes properly, in agreement with experiments [7], the optical Magnus effect in a circular waveguide. However, in the geometrical optics of Zel'dovich and Liberman, the equation for momentum is free of the correction that corresponds to the first equation in (27) and that is responsible for Berry phase. The matter of the fact that in the paper [7] the polarization of a wave corresponds to its independent degree of freedom, for which the evolutionary equations are written; this adds complexity to the theory. Meanwhile, as it has been shown, this is not the case. For every eigenmode, the polarization (right-hand or left-hand circular) is strictly fixed (the helicity is the adiabatic invariant of a photon), while the polarization evolution for an arbitrarily polarized wave is nothing but the result of the interference of two eigenmodes of fixed polarizations. This kind of interference is completely described within the context of our theory.

It follows from the above that our theory makes a prediction about a new phenomenon, which is not present in the theory of the optical Magnus effect. In papers [7,8], the *deviation of the ray center of gravity* in relation to the polarization has been described. For example, this deviation is zero for a linearly polarized ray. Meanwhile, as it has been shown, a single linearly polarized ray simply does not exist. When propagating, this ray will split into two circularly polarized independent rays. In Subsection 4.2, we suggest the simple scheme of the experiment for observing the predicted effect of splitting of a noncircularly polarized ray into two circularly polarized ones.

From Eqs. (26) and (27) along with Eqs. (17) the expressions for the group velocities of the waves of right-hand and left-hand polarizations follow:

$$\mathbf{v}_g^\pm = \frac{c}{n^{(0)}} \left( \mathbf{l} \pm \frac{1}{k_0 p^{(0)3}} [\dot{\mathbf{p}} \times \mathbf{p}]^{(0)} \right) = \frac{c}{n_0} \left( \mathbf{l} \pm \frac{1}{k_0 n_0} [\nabla \ln n_0 \times \mathbf{l}] \right) . \qquad (28)$$

The above formula points to the fact that the group velocities of the right-hand and left-hand waves are equal in magnitude in the given approximation: $\left|\mathbf{v}_g^+\right| = \left|\mathbf{v}_g^-\right| + O(k_0^{-2})$, and deflect in the opposite directions from the ray of zero-order approximation.

## 4. EXAMPLES: RAY SHIFTS IN CIRCULAR WAVEGUIDES

**4.1 Rays in the paraxial approximation.** In [7], the rotation of the plane of meridional right-hand and left-hand circularly polarized rays has been calculated in the mode approximation. Then, in [8], the same effect has been calculated from the suggested correction to the geometrical optics equations, which is similar to the second equation in (27). The results of these calculations are found to be coincident and in good agreement with experimental data [7]. Thus we may assert that the theory suggested above also describes adequately the optical Magnus effect in a circular waveguide. Nevertheless, we would like to present the calculation of this effect, which is based not on equations (27) but immediately on the initial equation (16), virtual ray trajectories, and the presence of Berry phases. This will allow us to demonstrate clearly the physical and geometrical meaning of the theory constructed above.

Consider a meridional ray propagating in the positive $z$ direction in a circular waveguide with a gradient parabolic profile in the paraxial approximation. Let the refractive index be the following function of the distance $r$ to the waveguide center

$$n(r) = n_0 \left[ 1 - \Delta \left( \frac{r}{r_0} \right)^2 \right], \qquad (29)$$

where $\Delta \ll 1$, while $n_0$ and $r_0$ are the characteristic refractive index and the radius of the waveguide. Here and further, unless otherwise specified, we imply the values of zero-order approximation Eqs. (17); for the sake of simplicity the indices are omitted. Let us introduce the natural cylindrical coordinates $(r, \varphi, z)$. The ray propagation process will be observed from the waveguide end (Fig. 1). As it follows from Eqs. (16) or (18), the ray displacement is proportional to the momentum gradient of its Berry phase per a unit of length. Although the meridional ray represents a plane curve and its Berry phase is zero, the adjacent, virtual, rays may possess the Berry phase, and hence, its gradient will be different from zero.

First note that variations in momentum components $p_r$ and $p_z$ do not move the trajectory away from the propagation plane, and hence, the derivatives $\partial/\partial p_r$ and $\partial/\partial p_z$ of Berry phase of the meridional ray are equal to zero. Thus, only $\varphi$-component of the momentum gradient of Berry phase of the meridional ray will be different from zero. Therefore, with the use of (18), the following equation for the desired ray shift can be written:

$$\frac{d\mathbf{r}^{(1)}}{ds} = \mp k_0^{-1} \frac{\partial}{\partial p_\varphi}\left(\frac{d\vartheta}{ds}\right)\mathbf{j}_\varphi . \qquad (30)$$

Here, $\mathbf{j}_\varphi$ is the unit vector directed along the $\varphi$-coordinate. As it was noted, the Berry phase of the meridional ray ($p_\varphi = 0$) equals zero. Consequently, to determine the gradient (30), we must consider a ray close to the meridional one and possessing small value of $p_\varphi \neq 0$. The ray trajectories (given by equations (17)) in parabolic profile (29) admit analytical solutions and is fully considered in [14]. It is well known [2,4–6] that Berry phase (11) of the ray is equal to the oppositely-signed area that is swept by the tangential vector $\mathbf{l}$ on a unit sphere. In the Appendix, it is shown that in the paraxial approximation the tangential vector traces an ellipse on a small section of the unit sphere. The area of this ellipse equals

$$S = \frac{\sqrt{2\Delta}\, r p_\varphi}{n_0 r_0} . \qquad (31)$$

Formula (31) with the opposite sign specifies the increment of Berry phase $\vartheta$ over one trajectory period $z_0 \approx \frac{\sqrt{2}\pi r_0}{\sqrt{\Delta}}$ (see (A2)). Therefore, the increment of Berry phase over a unit of length can be written (taking into account the sign) as

$$\frac{d\vartheta}{ds} \approx -\frac{S}{z_0} \approx -\frac{\Delta r p_\varphi}{\pi r_0^2 n_0} . \qquad (32)$$

Hence it follows that the correction to the refractive index for a spiral trajectory is

$$n^{(1)} \approx \mp \frac{\Delta r p_\varphi}{\pi k_0 r_0^2 p_z} . \qquad (33)$$

Formulas (32) and (33) are actually the averaging of the corresponding values over a period of the trajectory. It is quite sufficient, since ray shift is immeasurable for smaller scales; the effect shows itself over many periods. By substituting Eq. (32) into Eq. (30), we arrive at

$$\frac{d\mathbf{r}^{(1)}}{ds} \approx \pm \frac{\Delta r}{\pi k_0 r_0^2 n_0} \mathbf{j}_\varphi \ . \tag{34}$$

Since the shift $\mathbf{r}^{(1)}$ is proportional to $r$ and is directed along the $\varphi$ coordinate, it can be written as the shift in $\varphi$:

$$\frac{d\varphi^{(1)}}{ds} \approx \pm \frac{\Delta}{\pi k_0 r_0^2 n_0} = const \ . \tag{35}$$

Expression (35) indicates that *all* ray trajectories (regardless of $p_\varphi$) and not only the meridional ones are rotated uniformly clockwise or anticlockwise depending on the polarization sign (Fig. 1). This inference explains the good agreement between the mode approximation experiments [7] and the ray theory. The trajectory rotation angle is found from Eq. (35):

$$\varphi^{(1)} \approx \mp \frac{\Delta}{\pi k_0 r_0^2 n_0} z \ . \tag{36}$$

This formula corresponds exactly to the results obtained in [7]. Its derivation has revealed that the optical Magnus effect is indeed closely related to the presence of the Berry phase in the system and its anisotropy. Let us remark that if one considered the ray similar to the meridional one in a planar waveguide, the ray shift would not be observed. This is because Berry phase in a planar waveguide is *identically* equal to zero for all rays. At the same time, the initial meridional ray may have precisely the same trajectory as it has in a circular waveguide.

**4.2. Splitting of a circular ray.** Considering the ray shift effect from the viewpoint of the presence of Berry phase of the adjacent, virtual, rays, we can propose a straightforward scheme for observing both the optical Magnus effect and the ray *splitting*. Let us consider a finite ray propagating along a circle in the $z = const$ plane in a radially inhomogeneous medium (circular gradient waveguide) (Fig. 2). This kind of a ray corresponds to so-called modes of a whispering gallery. The ray by itself represents a plane trajectory with Berry phase equal to zero ($2\pi$, to be more specific). However, the adjacent rays with small $p_z \neq 0$ become spiral and gain a geometric phase. This suggests at once that the ray considered will shift in the direction of positive or negative $z$ according to its polarization (see Fig. 2). If both of the waveguide ends are open, the right-hand polarized wave will emerge from one end, whereas the left-hand polarized wave will emerge from the opposite end. This kind of experiment can be used to demonstrate the *splitting* of a ray of mixed polarization into two circularly polarized eigen rays. Indeed, if the initial ray is linearly polarized, the right and left circularly polarized radiation appears from two waveguide ends to the observer. Notice that, according to the interpretation of the Magnus effect given in [7,8], the linearly polarized ray is free from any displacement. In fact, these works estimate only the shift of the ray center of gravity and this shift is zero for a linearly polarized ray (since the shifts of two equal circularly polarized ray compensate each other). The splitting of a ray of mixed polarization into two circular rays can be obtained only from the proposed theory. Hence the experiment under discussion can support our theory.

The analyzed effect can be estimated easily by analogy with the above example. It is readily seen that the ray will be shifted in $z$ coordinate according the following equation:

$$\frac{dz^{(1)}}{ds} = \mp k_0^{-1} \frac{\partial}{\partial p_z}\left(\frac{d\vartheta}{ds}\right) . \qquad (37)$$

The tangent vector **l** of the initial ray is moving along the equator of the unit sphere, and hence the Berry phase over one period of the trajectory equals $2\pi$ (the unit hemisphere area). (At Fig. 2 we consider the initial ray that corresponds to the anti-clockwise movement when seen from the negative $z$ side. Therefore the area swept by the tangent vector on the unit sphere is negative and the Berry phase is positive.) For the ray with small $p_z$, the tangent vector will be moving along the parallel close to the equator; this will result in a small deviation of the geometric phase from $2\pi$. The parallel's latitude is $p_z/p \approx p_z/n_0$, and the Berry phase over one period equals

$$\vartheta \approx 2\pi - \frac{2\pi p_z}{n_0} . \qquad (38)$$

To obtain Berry phase gained by a wave over a unit of the trajectory length, expression (38) should be divided by the period length $2\pi r$:

$$\frac{d\vartheta}{ds} \approx -\frac{p_z}{n_0 r} \quad . \tag{39}$$

The term $2\pi$ in Eq. (38) has been omitted as inessential. By substituting Eq. (39) into Eq. (37) we have

$$\frac{dz^{(1)}}{ds} \approx \pm \frac{1}{n_0 k_0 r} = const \quad . \tag{40}$$

Equation (40) demonstrates the expected uniform displacement of the initial ray along $z$. In order to rewrite this displacement in an easy-to-use form, represent the trajectory length as $s = 2\pi r N$, where $N$ stands for the number of ray revolutions (periods). Then, upon integrating Eq. (40), we arrive at

$$z^{(1)} \approx \pm \frac{2\pi N}{n_0 k_0} = N\lambda \quad , \tag{41}$$

where $\lambda$ is the wavelength that corresponds to the refractive index $n_0$. Thus, with the characteristic length of the waveguide of $2L$, the circularly polarized ray has to complete $n_0 k_0 L / 2\pi = L/\lambda$ revolutions to leave the waveguide.

## 5. CONCLUSIONS

Above, the modified geometrical optics theory has been constructed for a smoothly inhomogeneous isotropic medium. In our derivations, we rely in large measure on the concept of locality and nonlocality, which allows us to find the proper way of separating complex phases and complex amplitudes of the wave solutions. It turns out that all nonlocal terms should be assigned to the wave phase and not to the amplitude. We have derived the first-order geometrical optics equations that properly and in a uniform way describe the Berry's geometric phases and the optical Magnus effect [4–6,7,8] (relationship between Berry's phase and Magnus effect was discussed also in paper [15]).

We have shown that in the first geometrical optics approximation a smoothly inhomogeneous locally isotropic medium becomes weakly anisotropic. The eigenmodes of this medium are the waves of right and left circular polarizations. This is due to the fact that the polarization form of circular waves remains unchanged during their propagation in a smoothly inhomogeneous medium. (An elliptically polarized wave changes its own polarization in accordance with the Rytov law [1–3], which is merely the result of the interference of two eigenmodes with different phase velocities.[3]) The eikonals of the right and left circular modes differ by the arising Berry phase of opposite signs Eq. (10). Hence, with the use of the eikonal equation, we have obtained the effective refractive indices (14) for circular modes. An essential dependence of the Berry phase not only on the coordinates but also on the wave vector direction determines a weak anisotropy of the medium.

From the Hamilton principle, for the obtained refractive indices, we have constructed ray equations (17) and (18), which involve the correction terms of the first order in $k_0^{-1}$. These corrections, being proportional to the spatial and momentum gradients of Berry phase, respectively, determine the deviations in momentums and coordinates for right and left circular waves. We have used the separation of local and nonlocal terms to bring these equations to a more convenient form like Eqs. (22), (26) or (27). At the same time, we have shown that the correction in the equation for momentum causes the difference in *absolute value* of the *phase* velocities, while the correction in the equation for coordinates is responsible for the difference in *direction* of the *group* velocities. The former effect describes the appearance of Berry phases of the wave solutions, whereas the latter one is associated with the deviation of the rays of different polarizations, which has been called before the optical Magnus effect [7,8].

---

[3] The conclusion about difference in phase velocities of right and left circular waves is contained already in the pioneer work [1]. However, the corresponding formula in that paper is not correct because the calculations were made in the rotational reference frame related to Frenet trihedron of the ray.

Hence the Berry phases, as well as the optical Magnus effect, are the accompanying phenomena that arise in the same order $k_0^{-1}$ in the geometrical optics equations. These phenomena describe the divergence in phase and trajectory, respectively, between the waves of different polarizations. We have found that the formula for the ray shifts for different polarizations Eq. (25) is geometric in character, just like the Berry phase, and represents a contour integral in the momentum space. Thus, both the optical Magnus effect and the Berry phase are *fundamental nonlocal topological phenomena*. It follows that in a one-dimensionally inhomogeneous medium (the medium with plane ray trajectories and free from Berry phases) the ray shift does not occur.

In addition to the above-listed findings, the suggested theory predicts a novel effect, which is not contained in the preceding theory of the optical Magnus effect [7,8]. Namely, a ray of mixed polarization not only undergoes the displacement of its center of gravity but also *splits* into two independent rays of right and left circular polarizations. Thus, in the approximation considered, *no* independent ray of arbitrarily mixed polarization exists. This ray may occur only as a result of the interference of the circular eigen rays propagating along different trajectories.

Our theory follows immediately from Maxwell's equations, the eikonal equations, and the Hamilton equations for rays. This theory describes from a unified standpoint repeatedly observed phenomena: Berry phase and the optical Magnus effect, which confirms its validity. Note also that the correction obtained in the coordinate equation of geometrical optics is exactly in line with the correction introduced by Liberman and Zel'dovich [8]. Consequently, this correction describes reliably the experimental data associated with the optical Magnus effect [7]. At the same time, geometrical optics of the paper [8] is free of the correction of the same order in the momentum equation (it is responsible for Berry phase), since in [8] evolution of the polarization is described by separate equation.

In parallel with the general theory, we have analyzed particular examples (both familiar and novel) of ray displacements for different polarizations. They fully support the inferences of our theory. With the help of the theory suggested, we have succeeded in calculating and analyzing the ray shifts associated with the optical Magnus effect. We have also proposed a novel scheme of the experiment that allows one to observe the splitting effect for the rays of mixed polarization.

It worth noticing that the effects of the ray deviations have the same order in magnitude, $k_0^{-1}$, as the ray diffraction. Therefore the diffraction spreading interferes significantly with the ray splitting. Nevertheless observations of the ray deviations are possible (see, for example, [7]) against the background of the diffraction spreading, since they are connected with the polarization characteristics of the ray.

Finally note that, owing to the general character of Berry phase as the phenomenon observed in dynamic systems, the analogs for the optical Magnus effect would be expected to occur in many systems. In particular, the effects of this kind occur during the propagation of quantum particles with a spin in external fields (see, for example, [16,17] and references there).

## ACKNOWLEDGMENT

The authors are grateful to Yu. A. Kravtsov and V. A. Permyakov for their interest to the work and fruitful discussions. The work was partially supported by INTAS (grant 03-55-1921).

# APPENDIX A: THE MOTION OF THE TANGENTIAL VECTOR OF A RAY IN A CIRCULAR WAVEGUIDE OF PARABOLIC PROFILE

Bellow we consider the ray equations in the geometrical optics zero approximation. It is readily seen from equations (17) (see also [3,14]) that in cylindrically inhomogeneous medium a wave possesses two ray invariants that are constant along the trajectory $I_1 = p_z$ and $I_2 = p_\varphi r / r_0$. Considering that $\mathbf{l} = \mathbf{p}/n$, these invariants can be rewritten in terms of the tangent vector component. Note also that in the paraxial approximation, the vector $\mathbf{l}$ is almost aligned with $z$-axis. The transversal component can be written as

$$l_\perp^2 = l_r^2 + l_\varphi^2 = 1 - l_z^2 = 1 - \frac{I_1^2}{n^2(r)} \approx 1 - \frac{I_1^2}{n_0^2} - 2\Delta\left(\frac{r}{r_0}\right)^2 . \tag{A1}$$

Here and further all calculations are performed in the first-order approximation in $\Delta \ll 1$. To derive the dependence of $l_\perp$ on the ray coordinate $s$ (which practically coincides with $z$ in the paraxial approximation), we should substitute the equation for the ray trajectory $r(z)$ into Eq. (A1). For the parabolic profile (29), the ray trajectory can be obtained analytically from Eqs. (17). Its projection onto a circular cross-section of the waveguide represents an ellipse (Fig. 1) and is given by equation [14]

$$r = \left[\frac{r_1^2 + r_2^2}{2} - \frac{r_1^2 - r_2^2}{2}\cos\left(\frac{z}{z_0}\right)\right]^{\frac{1}{2}}, \quad z_0 = \frac{\sqrt{2}\pi r_0 p_z}{\sqrt{\Delta}n_0} \approx \frac{\sqrt{2}\pi r_0}{\sqrt{\Delta}} . \tag{A2}$$

Here, $z_0$ is the period of the ray trajectory, while $r_1$ and $r_2$ are the major and minor ellipse semiaxes, which are equal to

$$r_{1,2}^2 = \frac{r_0^2}{4n_0^2\Delta}\left[\left(n_0^2 - I_1^2\right) \pm \sqrt{\left(n_0^2 - I_1^2\right)^2 - 8\Delta n_0^2 I_2^2}\right] . \tag{A3}$$

By substituting Eq. (A2) into Eq. (A1), we obtain

$$l_\perp^2 \approx 1 - \frac{I_1^2}{n_0^2} - \Delta\frac{r_1^2 + r_2^2}{r_0^2} + \Delta\frac{r_1^2 - r_2^2}{r_0^2}\cos\left(\frac{4\pi z}{z_0}\right) . \tag{A4}$$

Hence it follows that the end of the tangent vector is tracing an ellipse around the pole on the unit sphere. The pole on the sphere corresponds exactly to $z$-direction, while the ellipse occupies a small area, within which the surface may be treated as a part of a plane. Using Eq. (A3), we can derive from Eq. (A4) that the squares of the ellipse semiaxes are equal to

$$a^2 = 1 - \frac{I_1^2}{n_0^2} - 2\Delta \frac{r_2^2}{r_0^2} \approx 1 - \frac{I_1^2}{n_0^2} \ , \quad b^2 = 1 - \frac{I_1^2}{n_0^2} - 2\Delta \frac{r_1^2}{r_0^2} \approx \frac{2\Delta I_2^2}{n_0^2 - I_1^2} \ . \tag{A5}$$

The area of the ellipse is

$$S = \pi a b \approx \frac{\sqrt{2\Delta} I_2}{n_0} = \frac{\sqrt{2\Delta} r p_\varphi}{n_0 r_0} \ . \tag{A6}$$

The expression (A6) is obtained with the sign of the oriented area in mind: this sign will change with the sign of $p_\varphi$.

**FIGURE CAPTIONS**

**Fig. 1.** Propagation of almost meridional rays in a circular waveguide with a gradient profile (view from the negative z end). To the left: the meridional ray (thick line) and the virtual ray, close to the meridional one (dashed line) in the zero approximation of the geometrical optics. To the right: the shift of the left (blue line) and right (red line) polarization rays relative to the trajectory of the zero approximation (thick black line).

**Fig. 2.** Splitting of a finite ray of mixed polarization (thick black line) into two rays of left (blue line) and right circular polarization (red line) in a circular waveguide (angle view). The z axis is directed from left to right.

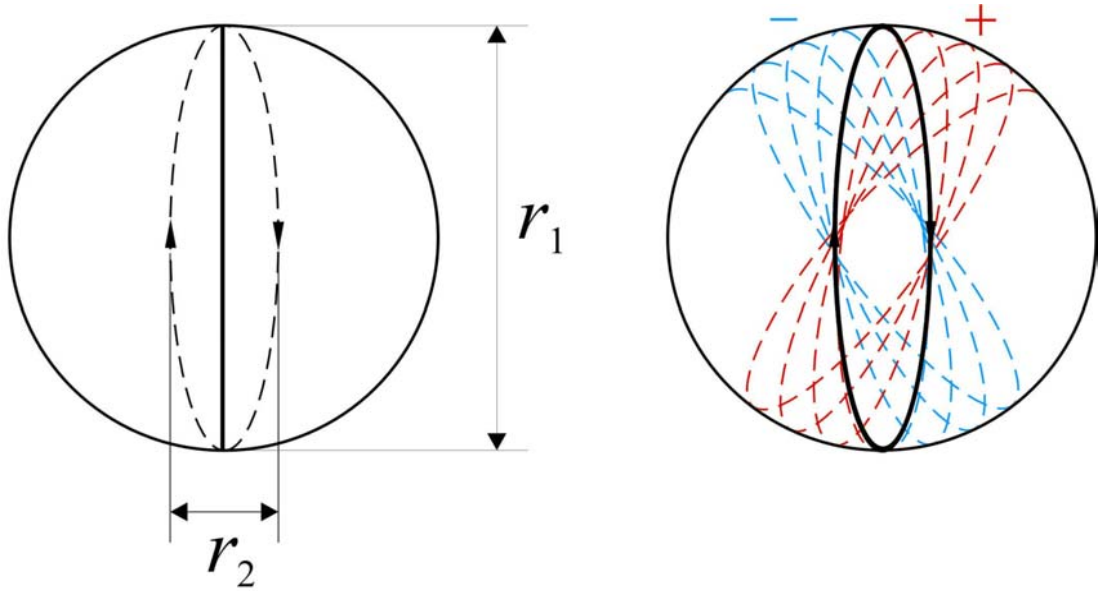

**Fig. 1**

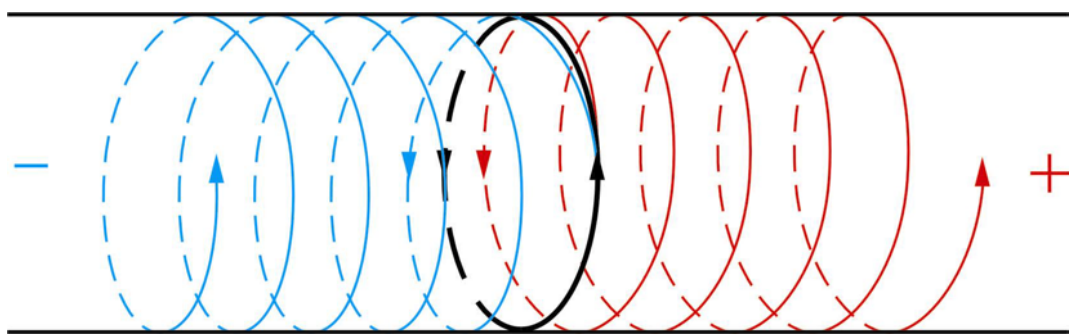

**Fig. 2**